\documentclass[12pt,letterpaper,twocolumn]{article} 

\usepackage{ol2}
\usepackage[draft]{hyperref}
\usepackage{amsmath}

\begin{document}
\twocolumn[ 

\title{Lineshape Asymmetry for joint CPT and three photon N resonances}

\author{Cindy Hancox,$^{1,2*}$, Michael Hohensee,$^{2**}$, Michael Crescimanno$^{3}$, David F. Phillips and Ronald L. Walsworth$^{2}$}
\address{$^1$Massachussetts General Hospital, Boston, Massachussetts, 02114}
\address{$^2$MS-59, Harvard-Smithsonian Center for Astrophysics, \\ 60 Garden St., Cambridge, Ma. 02138}
\address{$^3$Department of Physics and Astronomy, \\ Youngstown State University, Youngstown, OH 44555}
\address{$^*$Present Address: Massachussetts General Hospital, Boston, Massachussetts, 02114}
\address{$^{**}$Corresponding author: hohensee@fas.harvard.edu}

\begin{abstract}
We show that a characteristic 
two photon lineshape asymmetry arises 
in coherent population trapping
(\emph{CPT}) and three photon (\emph{N}) resonances because both 
resonances are simultaneously induced by 
modulation sidebands in the interrogating laser light. 
The \emph{N} resonance is a 
three-photon resonance in which a two-photon Raman 
excitation is combined with a resonant optical pumping field.
This joint \emph{CPT} and \emph{N} resonance can be 
the dominant source of lineshape distortion, 
with direct
relevance for the operation of miniaturized atomic frequency
standards.  
We present the results of both an experimental study and theoretical treatment of the asymmetry of the joint \emph{CPT} and \emph{N} resonance under conditions typical to the operation of an \emph{N} resonance clock.
\end{abstract}


\maketitle 
]

In compact, all-optical atomic clocks
employing Coherent Population Trapping (\emph{CPT})
\cite{Kitching00, Merimaa03, smallclock1, Vanier},
or three photon \emph{N} resonances\cite{3,6} 
interrogating light fields are
typically generated by current modulating a single-mode diode laser
with a large modulation index.  
This creates
additional optical fields. 
which can also drive atomic resonances.
As we show here both \emph{N} resonances \cite{3} and \emph{CPT}
resonances \cite{3,2}
are typically present. 
These two 
processes may then interfere and undergo
differential AC-Stark shifts, 
which lead generally to an asymmetric joint resonance.
Temporal variation in 
lineshape asymmetry has been shown
experimentally to lead to 
clock frequency instabilities\cite{8}. 

An \emph{N} resonance is a three-photon, two-optical-field absorptive
resonance. 
A probe field $\Omega_1$, resonant with the
transition between the higher-energy hyperfine level of the ground
electronic state and an electronically excited state, optically pumps
the atoms into the lower hyperfine level.
This probe field, $\Omega_1$, also acts on the lower hyperfine state
in combination with
a drive field
$\Omega_0$ detuned from the probe field by the atomic hyperfine
frequency $h$. Together, $\Omega_1$ and $\Omega_0$ create a
two-photon Raman resonance that coherently 
drives atoms from the lower
to the upper hyperfine level.
This causes increased absorption of
the probe field $\Omega_1$ in a narrow resonance with linewidth
$\Delta\nu$, set by the ground-state hyperfine decoherence rate. 

A practical \emph{N}-resonance clock 
creates the fields $\Omega_0$ and $\Omega_1$
using modulation of a single laser
source.  The laser carrier field $\Omega_0$ is detuned by 
approximately the ground state hyperfine frequency below the $F=2\rightarrow
F^\prime=1$ transition, and the modulation frequency is set so that the
first sideband $\Omega_1$ is resonant with 
that
transition, leading to an \emph{N}-resonance.
Additional sidebands such as $\Omega_2$ (see Fig.~1a) 
are also present. 
These sidebands participate with 
$\Omega_0$ and/or $\Omega_1$ in \emph{CPT} 
resonances,
which simultaneously compete with the 
\emph{N} resonances, leading to an overall lineshape asymmetry. 
While an ideal \emph{CPT} resonance produced by two optical fields will
not have an associated \emph{N} resonance, additional optical fields are
typically present in \emph{CPT} clocks driven by a modulated laser. At
the large modulation index used in \emph{CPT} clocks \cite{4}, higher order
sidebands which drive pairs of \emph{N} resonances are present.  For
pure phase modulation of the optical field the asymmetric signals
from these \emph{N} resonances cancel. In the presence of amplitude
modulation this cancellation is incomplete and residual asymmetry
may become significant.

Our experimental studies of joint \emph{N+CPT} resonances~\cite{4, 6, 10} used a beam of 795 nm
light from a diode laser, tuned near the $^{87}$Rb D$_1$
transition and modulated at the hyperfine frequency by an
electro-optic modulator (EOM). The laser light was
circularly polarized and sent through
a heated $^{87}$Rb vapor cell (65$^o$C,  Neon buffer gas of 30 Torr). 
The vapor cell was housed in high permeability
magnetic shields inside of which was a uniform longitudinal magnetic field (used
to split the Zeeman degeneracy) created by a solenoid. In results presented
here, the $\Delta m=0$, magnetic field-independent transition was studied.
A temperature stabilized Fabry-Perot cavity (FP) after the cell
selected the +1 sideband, whose intensity was measured with a photodiode.
In spectroscopy studies, the synthesizer driving the EOM
was locked to a hydrogen maser to provide high frequency stability.
Additionally, for elimination of the +2 sideband {\it before} the
cell, the laser beam was retroreflected off a second FP tuned to pass 
only the
+2 sideband. Two-photon lineshapes were
measured for various laser detunings and powers, along with the
presence or absence of the +2 sideband. In all data presented, the RF
(EOM) modulation index was fixed at 0.6.
For example, Figure 1b shows a typical measured lineshape 
for the transmitted probe field $\Omega_1$, illustrating the
asymmetry of the joint \emph{N+CPT} resonance.

To the usual three-state $\Lambda$-system used to 
model \emph{CPT}, we append an additional excited state 
with dipole coupling to the two ground states. 
This fourth state is assumed to  
be far off resonance and   
accounts for the non-resonant dipole polarizability
to which many excited states may contribute. The optical fields 
$\Omega_1$ and $\Omega_2$ form the \emph{CPT} system; whereas 
$\Omega_0$ and $\Omega_1$ 
participate in the \emph{N} resonance 
(see Fig.~1a). $\Delta$ is the (one photon) laser detuning 
of $\Omega_1$ and $\delta$ is the two-photon detuning
of the laser fields for both the \emph{CPT} and \emph{N} resonances. 

Using our model we find that 
to leading order in the optical fields, the transmission of the probe
field  $\Omega_1$ is proportional to $T(\Delta) - Im\bigl( \rho_{ab}
{{\Omega_0}\over{\Omega_1(h-\Delta)}}\bigr)$.  
Here
$T(\Delta)$ is the transmission
independent of the \emph{CPT} and \emph{N} resonances, 
$h$ is the hyperfine frequency,
and $\rho_{ab}$ is the ground state coherence. 
In steady state we find, 
\begin{align} 
\biggl(\Gamma-i(\delta+
&{{~|\Omega_1|^2-|\Omega_0|^2}\over{4(h-\Delta)}})\biggr)\rho_{ab} =
\nonumber \\ 
&  -i{{\Omega_1}\over{2}}\rho_{cb} + i{{\Omega_2}\over{2}}\rho_{ac}
\nonumber \\
&  + {{\Omega_1\Omega_0}\over{4(h-\Delta)}}(\rho_{bb}-\rho_{aa}),
\end{align}
%
%
%
where $\Gamma$ is the ground state depolarization rate
and the subscripts refer to the atomic levels 
as shown in Figure 1a.  The second line in Eq. (1) is associated with the
\emph{CPT} resonance and the last line is the contribution of the 
\emph{N} resonance.
This expression is derived by adiabatically eliminating the
non-resonant states contributing to the atomic polarizability. The
equation for the ground state
population difference, $\rho_{bb}-\rho_{aa}$, is
structurally equivalent to Eq. (1), with an \emph{N} resonance term
proportional to $\rho_{ab}$ and a \emph{CPT} term linear in $\rho_{ac}$ and
$\rho_{bc}$. Since atomic coherences ({\it e.g.}, 
$\rho_{cb}$) scale as $\Omega/\Delta$,
the leading order \emph{CPT}  and \emph{N} resonance driving 
terms are of the
same order at relevant one-photon detunings.
Note that for the numerically calculated results presented below, 
the contribution from the full excited state manifold is utilized, 
whereas Eq. (1) is for only a single resonant excited state.


Numerical calculations of the probe field $\Omega_1$ 
transmission intensity are shown graphically in Figure 2 in the
approximation that the vapor cell is optically thin. 
The limiting cases of 
pure \emph{CPT} resonance
($\Omega_0 = 0$) and ideal 
\emph{N} resonance 
($\Omega_2 = 0$) 
are not centered at the same two-photon detuning because they 
experience different AC Stark shifts.
The relative amplitudes for \emph{CPT} and \emph{N}
resonances are typical of experimental results.

For comparison between theory and experiment we quantify the lineshape
asymmetry by fitting to a skew Lorentzian\cite{11}:
\begin{align} 
I(\delta) = C+ D\delta +{{A\Gamma+B(\delta-\delta_0)}\over{(\delta-\delta_0)^2 + \Gamma^2}} .
\end{align}
Here 
$I(\delta)$ is the transmitted $\Omega_1$ 
intensity of the combined \emph{N+CPT} resonance,
$\delta$ is the two-photon frequency, and  $A$, $B$, $C$, and $D$ are
fitting amplitudes. The fitting parameter $\Gamma$ is proportional to
the resonance linewidth, and $\delta_0$ is the two-photon resonance frequency.
The amplitudes $A$ and $B$ describe the symmetric (Lorentzian) and
antisymmetric (dispersive) components of the lineshape, respectively.  
We define the
line asymmetry as the dimensionless ratio $B/A$. As an example, 
consider the effect of non-resonant light fields:
these fields are time-varying in the rotating frame
and thus contribute AC Stark shifts to each transition.  The
sign of these shifts depends on the sign of the non-resonant
field's detuning. 
Different AC Stark  shifts for the \emph{CPT} and 
\emph{N} resonances force 
the maximum of the optical response away from 
$\delta = \delta_0$, 
creating an overall asymmetric lineshape. 
The $B$ term, linear in the two-photon
detuning $\delta-\delta_0$, accommodates this effect. 

We fit the skew Lorentzian (Eq. (2)) to both 
experimental data and 
numerical calculations using the 
model described above with atomic parameters for 
the $^{87}$Rb atom. We find that the
lineshape asymmetry $B/A$ decreases as overall laser intensity increases
in both cases (Fig.~3). 
This reduction of asymmetry 
with increasing 
optical power is
consistent with our model of the different 
multi-photon (nonlinear) 
natures of the \emph{CPT} and \emph{N} resonances.
As a
three-photon process, the \emph{N} resonance
contrast scales faster with laser power than that of the
two-photon CPT process (see Eq. (1)); 
thus lineshape asymmetry
is less pronounced
at large total power. 
Furthermore, the line asymmetry is greatly reduced over the entire range
when the +2 sideband is suppressed (in our experiments
by approximately 85\% in intensity)
{\it before} the light enters the atomic
vapor cell. Our model indicates that reduction of the +2 sideband 
greatly inhibits the \emph{CPT} 
resonance. 
The asymmetry $B/A$ in the two 
photon line shape is proportional to the ratio of the rabi 
frequencies $\Omega_2/\Omega_0$ and decreases with increasing 
ground state population difference. This connection 
between the parameterization of Eq. (2) and the full model 
explains the qualitative features
of Figures 3, 4.

An another example, our model explains the difference between the 
observed 
lineshape asymmetries of \emph{N+CPT} resonances
on the D$_1$ and D$_2$ optical transitions~\cite{10}.
Under identical conditions the 
\emph{CPT} resonance has lower contrast on the D$_2$ transition 
as compared to that of D$_1$~\cite{12},
This is 
a consequence
of direct depolarization transitions ($F=2 \rightarrow F'=3$) for the 
D$_2$ drive and 
the resultant suppression of optical pumping of the ground state. 
Under identical conditions, the D$_2$ ~\emph{N} resonance 
is more symmetrical and has higher
contrast than the D$_1$ 
\emph{N} resonance (Fig.~4).

In conclusion, we have shown that a  characteristic two-photon 
lineshape 
asymmetry arises in \emph{CPT} and \emph{N} resonances 
due to modulation sidebands in the interrogating laser light. 
A simple model for the 
combined effect
of these optical fields 
in the joint {\it N+CPT} system explains quantitatively many 
observed 
features of the lineshape asymmetry. 
Temporal variation in lineshape asymmetry 
contributes to clock frequency instability. The
effect described here is most relevant 
for \emph{N} resonance-based clock stability, but can contribute 
also to the optical response of \emph{CPT}-based clocks as well
when driving fields are not perfectly balanced. 
More experimental studies are underway to fully assess the
challenge of using a significantly distorted lineshape in a 
clock. This asymmetry is generic to
\emph{N} resonance-based clocks using current-modulated laser diodes.

We acknowledge support from ONR, ITAMP and the  Radcliffe Institute for 
Advanced Studies.



\par 
\vfill
\eject

\vskip .3in

\begin{figure}[htb]
\centerline{\includegraphics[width=9cm]{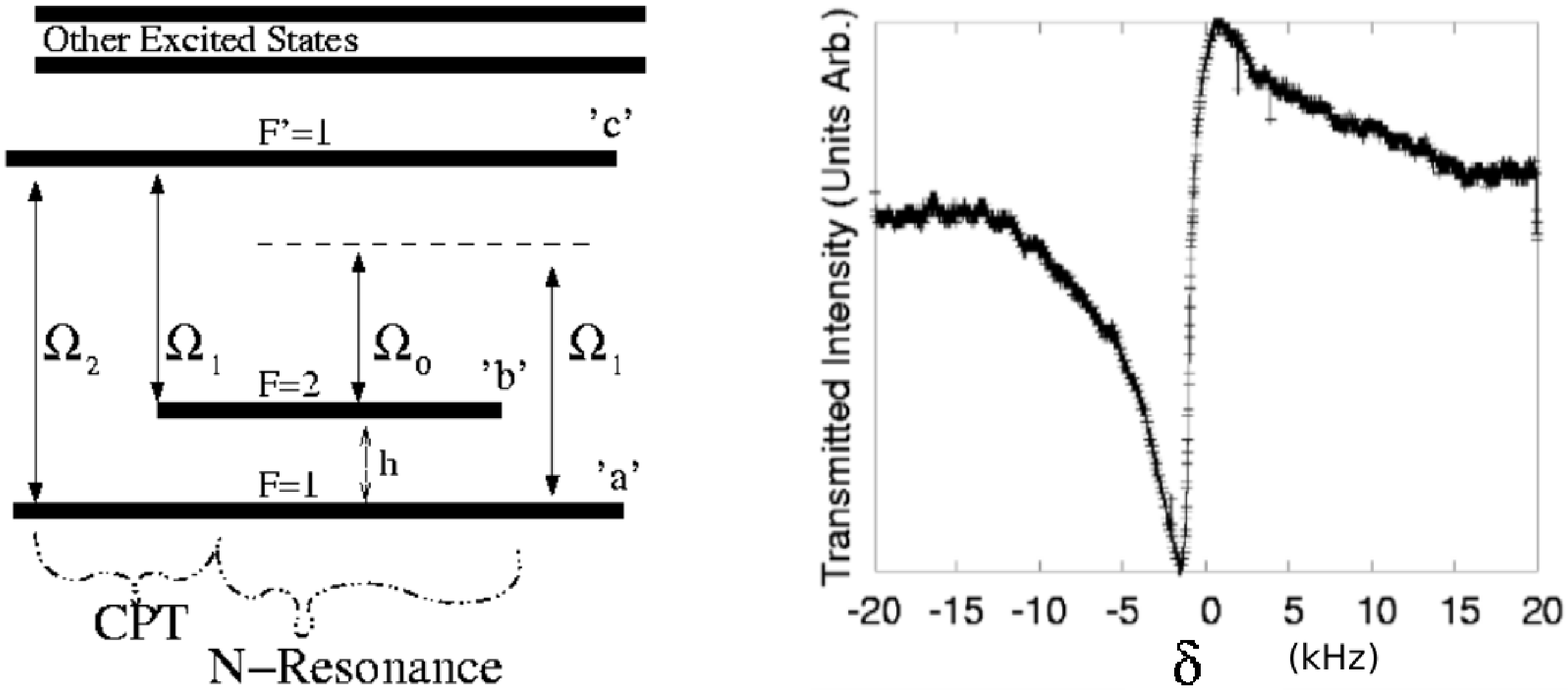}} 
\end{figure}

%

\noindent {\bf Fig. 1:} 

(a) Simplified level diagram with applied fields $\Omega_0$ 
(carrier) and $\Omega_1$, $\Omega_2$ (sidebands). 
~(b) Example experimental \emph{N+CPT} joint resonance, 
illustrating the typical asymmetry of the 
transmitted probe field $\Omega_1$ lineshape 
in the presence of the $\Omega_2$ sideband. 
The x-axis is the two-photon detuning. The laser 
power is .088 mW and the one photon detuning is about 350 MHz

\vskip .3in 

\begin{figure}[htb]
\centerline{\includegraphics[width=7cm]{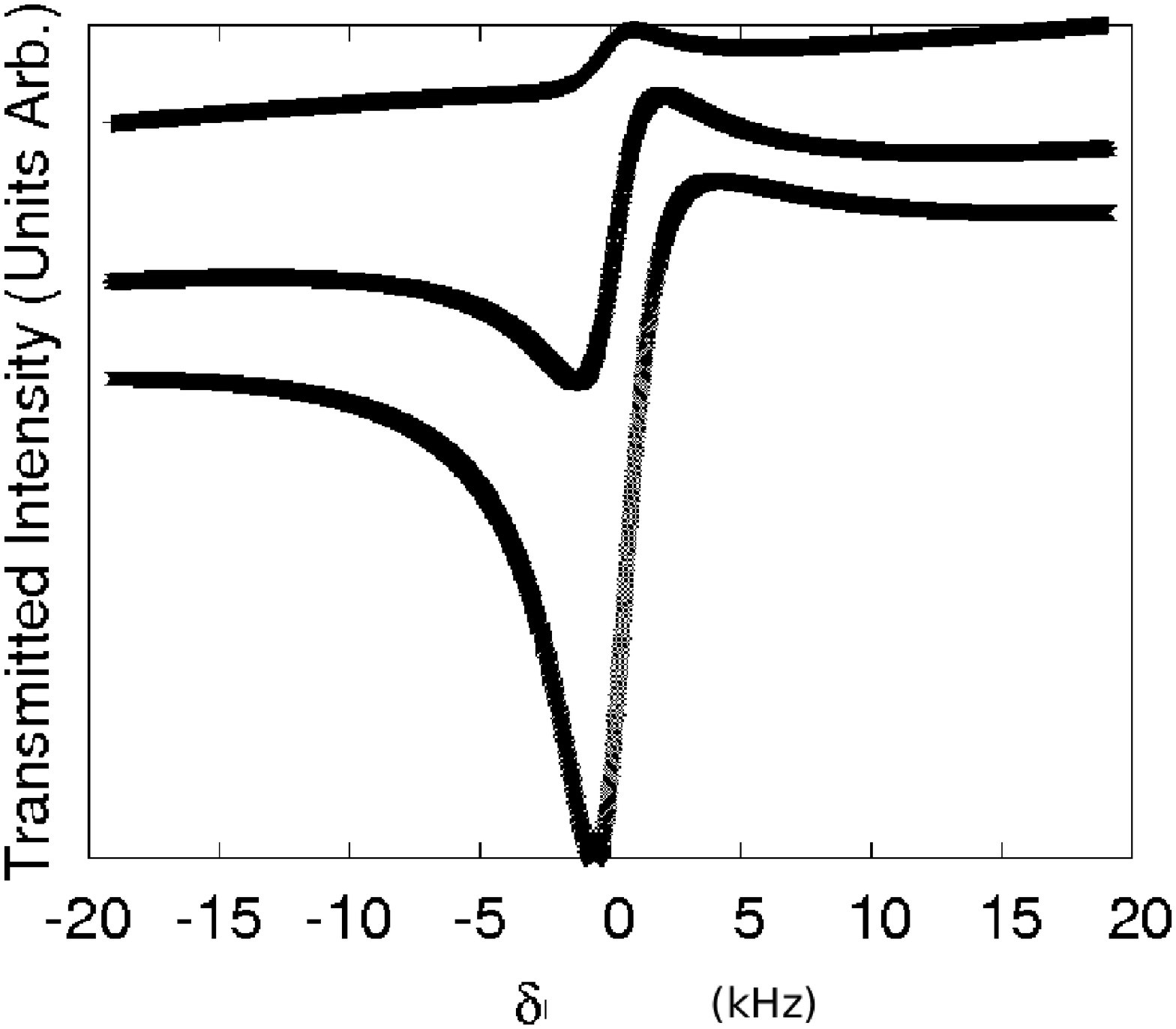}}
\end{figure}

\noindent {\bf Fig. 2:} 

Numerical calculations of the probe field $\Omega_1$ transmission 
intensity for  \emph{CPT} (top), \emph{N} (bottom) 
and joint \emph{N+CPT} resonances (middle), vertically offset 
for clarity. All parameters are for the $^{87}$Rb atom with 
$\Omega_0$ = 1 MHz for the middle and bottom traces and 
$\Omega_1$ = 100 kHz for all traces and $\Omega_2$ = 
6 kHz for the top and middle traces. 
The one-photon detuning used in these calculations is 100 MHz and so the 
background transmitted signal 
$T(\Delta)$ is not level. The  x-axis is two-photon 
detuning.

\vskip .3in 

\begin{figure}[htb]
\centerline{\includegraphics[width=7cm]{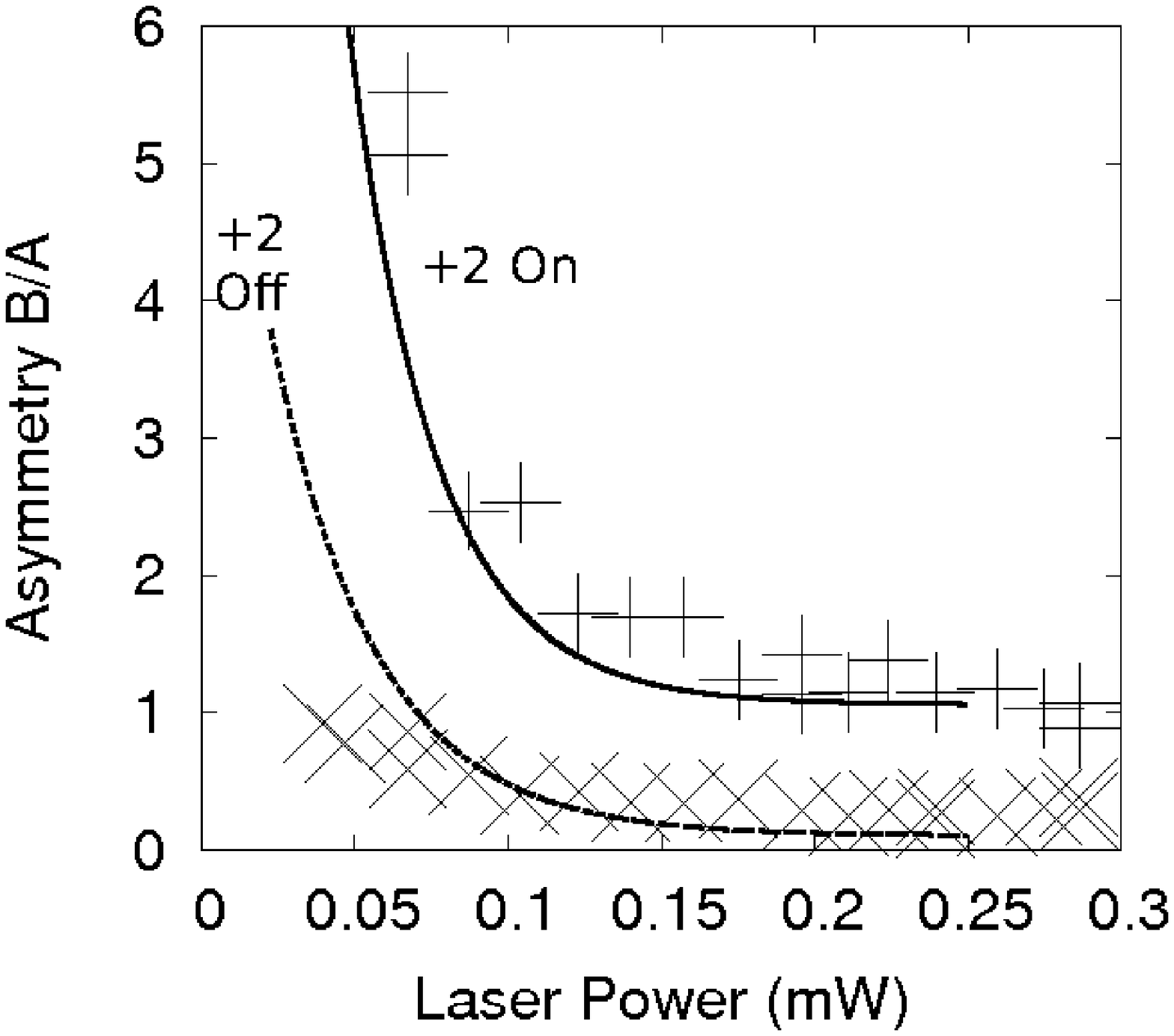}}
\end{figure}

\noindent {\bf Fig. 3:}

Lineshape asymmetry $B/A$ for \emph{N+CPT} resonances 
on the D$_1$ transition of 
$^{87}$Rb with and 
largely without (85\% reduced intensity) the 
+2 sideband. Numerical calculations based on our model are the lines, 
the ``+'' and ``x'' are the associated fitted $B/A$ from experimental data.

\vskip .3in

\begin{figure}[htb]
\centerline{\includegraphics[width=9cm]{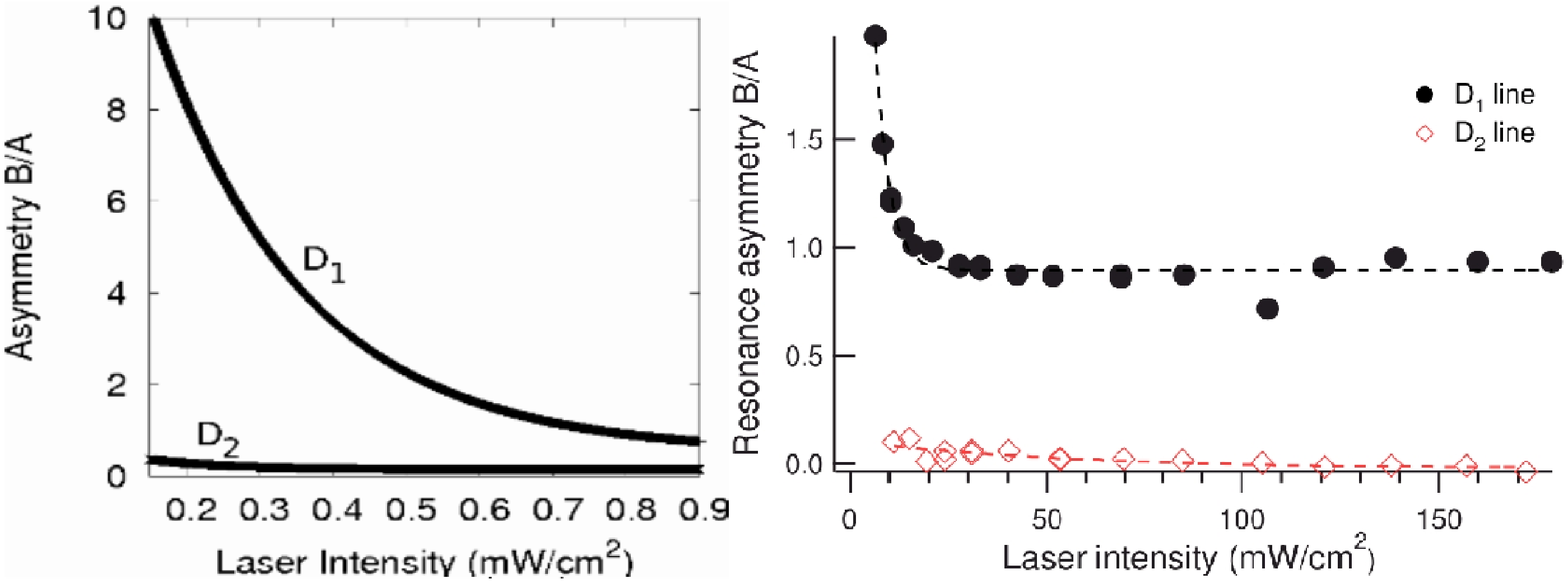}}
\end{figure}

\noindent {\bf Fig. 4:} 

Comparison of \emph{N+CPT} lineshape asymmetry 
for D$_1$ and D$_2$ transitions:  
(a) Numerical calculations and 
(b) experiment (data from \cite{10}).
Difference with 
Figure 3 is due to the 
different experimental parameters (buffer gas pressure, laser intensities 
and detunings) of \cite{10} that we have used in these
calculations. The range of intensities for which there is 
experimental data is not directly acessible to our theory model, but both 
model and experiment indicate saturation of the two-photon optical pumping
at high power. 


\begin{thebibliography}{99}

\bibitem{Kitching00} J.\ Kitching, S.\  Knappe, N.\  Vuki\u{c}evi\`{c}, 
L.\  Hollberg, R.\  Wynands and W.\  Weidmann, 
``A microwave frequency reference
based on VCSEL-driven dark line resonances in Cs vapor,'' 
IEEE Trans.\ Instr.\ Meas.\ \textbf{49}, 1313--1317 (2000).

\bibitem{Merimaa03} M.\ Merimaa, T.\  Lindvall, I.\  Tittonen, 
and E.\  Ikonen,
``All-optical atomic clock based on coherent population trapping in ${}^{85}$Rb,'' 
J.\ Opt.\ Soc.\ Am.\ B \textbf{20}, 273--279
(2003).

\bibitem{smallclock1} L. A.\  Liew, S.\ Knappe, J.\ Moreland, 
H.\ Robinson, L.\ Hollberg and J.\ Kitching, 
``Microfabricated alkali atom vapor cells,'' 
{\it Appl. Phys. Lett.} {\bf 84} \#14, 2694 
(2004). 

\bibitem{Vanier} J.\ Vanier, 
``Atomic clocks based on coherent population trapping: a review,'' 
{\it Appl. Phys. B} {\bf 81} 421 (2005). 


\bibitem{7} J.\ Kitching, H. G.\  Robinson, L.\ Hollberg, S.\ Knappe, 
and R.\ Wynands,
``Optical-pumping noise in laser-pumped, all-optical microwave frequency references,'' 
J. Opt. Soc. Am. B {\bf 18} 1676-1684 (2001). 

\bibitem{3} A. S.\ Zibrov, C. Y.\ Ye, Y. V.\ Rostovtsev, A. B.\ Matsko, 
and M. O.\ Scully,
``Observation of a three-photon electromagnetically induced transparency in hot atomic vapor,'' 
Phys. Rev. A {\bf 65}, 043817 (2002).

\bibitem{2} E.\ Arimondo, 
``Coherent population trapping in laser spectroscopy,''
Prog. Opt.  {\bf XXXV},  257  (1996).

\bibitem{8} D. F.\ Phillips, I.\ Novikova, C. Y.-T.\ Wang, R. L.\ Walsworth, and M.\ Crescimanno,
``Modulation induced frequency shifts in a coherent-population-trapping-based atomic clock,''
J. Opt. Soc. Amer. B, {\bf 22}, 305-310 (2005). 

\bibitem{4} I. \ Novikova, D. F.\ Phillips, A. S.\ Zibrov, R. L.\ Walsworth, A. V.\ Taichenachev, and V. I.\ Yudin,
``Cancellation of light shifts in an N-resonance clock,'' 
Opt. Lett. {\bf 31}, 622-624 (2006). 

\bibitem{6} S. \ Zibrov, I.\ Novikova, D. F.\ Phillips,A. V.\ Taichenachev, V. I.\ Yudin, R. L.\ Walsworth and A. S.\ Zibrov,
``Three-photon-absorption resonance for all-optical atomic clocks,'' 
Phys. Rev. A {\bf 72}, 011801 (2005).  

\bibitem{10} I. Novikova, D. F.\ Phillips, A. S.\ Zibrov, R. L.\ Walsworth, A. V.\ Taichenachev, and V. I.\ Yudin,
``Comparison of $^{87}$Rb N-resonances for 
D$_1$ and D$_2$ transitions," 
Opt. Lett. {\bf 31}, 2353-2355 (2006). 

\bibitem{11} S.\ Knappe  M.\ St\"ahler, C.\ Affolderbach, A. V.\ Taichenachev, V. I.\ Yudin and R.\ Wynands, 
``Simple parameterization of dark-resonance line shapes,'' 
Appl. Phys. B {\bf 76}, 57-63 (2003). 

\bibitem{12} M. St\"ahler, R.\ Wynands, S.\ Knappe, J.\ Kitching, L.\ Hollberg, A. V.\ Taichenachev, and V. I.\ Yudin,
``Coherent population trapping resonances in thermal $^{85}$Rb vapor; D$_1$ versus D$_2$ line excitation,'' 
Opt. Lett. {\bf 27}, 1472 (2002).

\end{thebibliography}
\end{document}